\documentclass[prd,aps,nofootinbib,preprintnumbers]
 {revtex4}
\scrollmode
\usepackage{amsmath}
\usepackage{amssymb}
\usepackage{eucal}
\numberwithin{equation}{section}
\newcommand{\ben}{\begin{eqnarray}}
\newcommand{\een}{\end{eqnarray}}
\newcommand{\la}{\label}

\begin{document}

 \title[The $\mathbf{(2+1)}$-dim Axial Universes and Dimensional Reduction]{The $\mathbf{(2+1)}$-dim Axial Universes --
 Solutions to the Einstein Equations,
 Dimensional Reduction Points, and Klein-Fock-Gordon Waves}

 \author{P. P. Fiziev\\
 Dept. Theoretical Physics, Sofia University
 ``St. Kliment Ohridski", \\ 5 James Bourchier
 Blvd., 1164 Sofia, Bulgaria\\ and \\ BLTP,
 JINR, Dubna, 141980 Moscow Region, Russia \\ and \\
 D. V. Shirkov, \\ BLTP, JINR, Dubna, 141980 Russia}

  \begin{abstract}
 The paper presents a generalization and further development of
 our recent publications where solutions of the Klein-Fock-Gordon
 equation defined on a few particular $D=(2+1)$-dim static
 space-time manifolds were considered. The latter involve toy
 models of 2-dim spaces with axial symmetry, including dimension
 reduction to the 1-dim space as a singular limiting case.

 Here the non-static models of space geometry with axial
 symmetry are under consideration. To make these models closer
 to physical reality, we define a set of ``admissible" shape
 functions $\rho(t,z)$ as the $(2+1)$-dim Einstein equations
 solutions in the vacuum space-time, in the presence of the
 $\Lambda$-term, and for the space-time filled with the standard
 "dust". It is curious that in the last case the Einstein
 equations reduce to the well-known Monge-Amp\`{e}re equation,
 thus enabling one to obtain the general solution of the Cauchy
 problem, as well as a set of other specific solutions involving
 one arbitrary function. A few explicit solutions of the
 Klein-Fock-Gordon equation in this set are given.

 An interesting qualitative feature of these solutions relates to
 the dimension reduction points, their classification, and time
 behavior. In particular, these new entities could provide us
 with novel insight into the nature of P- and T-violation, and
 of Big Bang. A short comparison with other attempts to utilize
 dimensional reduction of the space-time is given.

\end{abstract}

\sloppy
\maketitle
 \section{Introduction}
  The idea of reducing the number of {\em topological} dimensions of
 the physical space at small distances (proposed recently \cite{Shir10})
 was implemented in our previous papers \cite{DVPF10,PFDV10} for the
 $(2+1)$-dim space-times
 which comprise the $2$-dim static axial spaces with an arbitrary
 shape function $\rho(z)\geq 0\,.$ This was done to develop general
 methods and get insight into possible features of physics in such
 a specific variable geometry, including dimensional reduction (DR).

In fact, dimensional reduction of the physical space in general relativity (GR) can be regarded as an unrealized
and as yet untapped consequence of Einstein's equations (EEqs)
themselves which takes place around singular points of their solutions.
The oldest indication of this still not studied phenomenon could be found as early as in
the well known 1921 Kasner solution of EEqs \cite{Kasner}, \cite{Landau}.
For a short history of the Kasner solution and its modern applications see \cite{Wainwright}.
It is well known that if we consider, for example, a $3$-dim space-cube, according to the Kasner solution
there are possible two types of its evolution, approaching the singularity:

1. Pancake type evolution: In this case one of the dimensions tends to zero and the cube becomes a $2$-dimensional
square.

2. Cigar type evolution: In this case two of the dimensions tend to zero and the cube becomes a $1$-dimensional line.

Thus, the Kasner solution demonstrates a clear trend toward dimensional reduction,
but the evolution {\em after} the singular point has never been considered.
The situation resembles the one in the two-body collision problem in the  Newtonian
gravity  before the invention of the Levi-Cevita continuation of the solutions after the collision.
In the present paper we consider the dimensional reduction as a dynamical problem for EEqs.

 To begin with, a more general time dependent axial geometry
 of the $2$-dim space is analyzed. The $\{t,z\}$-dependence
 of the shape function $\rho(t,z)\geq 0$ is obtained by solving the
 EEqs in $(2+1)$-dim space-times with axial
 spaces. For brevity, we refer to these specific space-times as to
 the $(2+1)$-dim {\em axial universes} (AxU). It turns out that
 despite the fact that the EEqs fix quite firmly the axial geometry
 under consideration, there still remains a variety of dynamically
 admissible space-time manifolds, including some of the
 previously studied static ones.

 It is well known that in any $(2+1)$-dim GR universe the {\em local}
 degrees of freedom, which may be related with gravitational waves,
 are freezed and we have no freely moving excitations of the
 gravitational field in space-times with trivial topology.
 There exist quite large literature on $(2+1)$-dim GR, see, for
 example \cite{Carlip1}, \cite{Carlip2}, and several hundred references therein.
 Unfortunately, there one cannot find consideration of our AxU which are specific
 solutions, not related with the $(2+1)$-dim models of quantum gravity studied in the literature
 or with $(2+1)$-dim black holes
 - the main subject and motivation for the previous investigations of the $(2+1)$-dim GR.
 As a result, we can extract from the existing literature only some results of general character:

 If the fundamental group of the space-time is nontrivial, a finite number
 of {\em global} gravitational degrees of freedom remains and
 provides the classical basis for quantum theory of gravity. While
 this feature makes the theory simple, it does not
 quite make it trivial. For example, it simplifies drastically the
 analysis of the dynamics described by the EEqs in the $(2+1)$-dim
 universes. In the case of AxUs the axial symmetry yields an additional
 simplification of the physical problem and ensures the existence of
 nontrivial fundamental group. As we shall see, it turns even possible
 to find the general solution of the EEqs for different matter contents
 of the $(2+1)$-dim AxUs \eqref{ds2_1_2} and to study the novel physical
 phenomena related with the variable topological dimension in them,
 which is the main subject of the present article.

Another general result in $(2+1)$-dim GR is the Birkhoff-like theorems \cite{Visser09, Zanelli}
which may be viewed as a no hair theorem for stars in (2+1) dimensions:
The exterior geometry of an axially symmetric star is completely specified by the mass, angular momentum,
and cosmological constant. Those results are only partially related with our topic, but
some of the used in the cited articles general considerations may be also valuable for us.

 The solutions of the Klein-Fock-Gordon equation (KFGEq) \cite{Klein,Fock_a,Fock_b,Gordon}
 in the $(2+1)$-dim
 AxUs, which are consistent with the EEqs, are studied. One can
 consider the related field excitation as test particles, since we
 ignore the back-reaction of these excitations on the metric.
 Thus,
 we reach the usual natural separation of macro- and micro-physics.
 Indeed, in the real world the geometry of the observable Universe,
 governed by the EEqs is determined by about $10^{80}$
 protons (the Eddington number) and the same number of electrons
 \cite{JB}. From a physical point of view the influence of particles
 and fields, which we use for Earth-laboratory and Space experiments
 on this geometry, is obviously negligible. Hence, to obtain useful
 information for our domestic experiments, it is natural to study
 the behavior of test particles and fields on some space-time
 background defined by the solutions of the EEqs with
 some nonzero energy-momentum tensor in the rhs which describes a
 bulk of matter filling the Universe.

 \section{The $\mathbf{(2+1)}$-dim Time Dependent Axial Universes}
 Consider auxiliary flat Minkowski $(3+1)$-dim space-time
 $\mathbb{E}^{(1,3)}_{x^0 x^1 x^2 x^3}$ with the interval
 \ben\la{dsigma2}
 d\sigma^2=(dx^0)^2-(dx^1)^2-(dx^2)^2-(dx^3)^2.\een

 We introduce {\em axial universe} (AxU)
 as a hypersurface in $\mathbb{E}^{(1,3)}_{x^0 x^1 x^2 x^3}$
 defined by the equations:
 \ben \la{M}                                               
 \mathbb{M}^{(1,2)}_{t \phi z}: \left\{\begin{array}{cccc}
 x^0=t,\quad x^1=\rho(t,z)\cos\phi,\cr
 x^3=z,\quad x^2=\rho(t,z)\sin\phi,
 \end{array} \right. \een
 assuming $t\in(-\infty,\infty)$, $z\in(-\infty,\infty)$, and
 $\phi\in [0,2\pi]$.
 This pseudo-Riemannian $(2+1)$-dim manifold has a structure
 $\mathbb{M}^{(1,2)}_{t \phi z}=\mathbb{R}^{(1)}_t\otimes
 \mathbb{R}^{(1)}_z\otimes\mathbb{S}^{(1)}_{\phi}$, with
 $\mathbb{S}^{(1)}_{\phi}$ being a circle. Clearly, the space
 $\mathbb{M}^{(2)}_{\phi z}=\mathbb{R}^{(1)}_z\otimes
 \mathbb{S}^{(1)}_{\phi}$ is a $2$-dim rotational surface with
 time-dependent variable shape function $\rho(t,z)\geq 0\,.$
 Thus, for the AxU
 \ben\la{ds2_1_2}
 ds^2=\left(1 -\dot{\rho}^2\right)dt^2 - 2\dot{\rho}\rho^\prime dt
  dz-\left(1+{\rho^\prime}^2\right)dz^2-\rho^2 d \phi^2;\\
 \dot{\rho}=\partial_t\rho(t,z),\quad \rho^\prime=\partial_z\rho(t,z).\nonumber
 \een
 Further on, we consider the pseudo-Riemannian space-times
 \eqref{ds2_1_2}, with restriction  on the lapse
 function $\left(1 -\dot{\rho}^2\right)> 0$.
 This condition is needed to preserve the relativistic
 causality and the physical meaning of the time variable $t$.
 It ensures that the time variations of the shape of the axial space
 are not able to spread faster than the light, or with velocity of
 light.

Note that the most general metric with axial symmetry for $(2+1)$-dim GR is a starting point in the articles \cite{Visser09, Zanelli}.
Under proper choice of coordinates the authors perform a detailed  analysis of {\em rotating} axially symmetric
$(2+1)$-dim vacuum space-times and derive the Birkhoff-like theorems.
At the end they arrive at the stationary but not static rotating
Banados-Teitelboim-Zanelli vacuum metric for the exterior domain of a $(2+1)$-dim rotating star.

The AxUs with metric \eqref{ds2_1_2} presents completely different space-times.
Being neither stationary nor static, they describe {\em non-rotating} universes,
both vacuum ones (without or with $\Lambda$ term), and filled with matter.
As we shall see in the next sections,
the metric \eqref{ds2_1_2} produces  quite different form of the EEqs
and yields essentially  different physical solutions which are interesting
for study of the dimensional reduction:

 At the points where $\rho(t,z)=0$ the dimension of the $2$-dim
 axial space reduces. We call these points \ {\em the dimensional
 reduction points} \ (DRPs). In general, they move along the
 $z$-axis, i.e., for DRP we have $z=z^{drp}(t)$. There exist two
 possibilities:
\begin{enumerate}\label{DRP}
 \item The DRP may be isolated, i.e, in the small enough vicinity there are no other DRPs,
 for example, the vertex of a cone, without extension (symbolically: "$>\hskip -.18truecm \bullet$"),
 or the vertex of a two-sided cone (symbolically: "$>\hskip -.18truecm \bullet\hskip -.18truecm <$").

 \item The DRP is not isolated, i.e., there exist some part of the continuous $1$-dim straight line of
DRPs which passes through this point, for example, the vertex of a cone extended by part of a straight
line (symbolically: "$>\hskip -.18truecm \bullet\hskip -.1truecm \text{---}$")\footnote{In more
general geometries without axial symmetry the $1$-dim space can be part of a curved line.}.
\end{enumerate}

\section{The Einstein Equations for the Time Dependent $\mathbf{(2+1)}$-dim AxU}

 The nontrivial EEqs in the presence of matter with the
 energy-momentum tensor \ $T^i_j$, $i,j=0,1,2$ \ read
\begin{subequations}\la{Enstn:a,b}
\ben \hskip -2.6truecm\left.
 \begin{array}{ll}
 G^{.t}_{t.}\!= -
 \Big( \left(1-\dot\rho^2\right)\rho^{\prime\prime}+
 \dot\rho\rho^\prime\dot\rho^\prime\Big)
 \big/{{\rho\mathfrak{g}^2}}\!=\!
 T^{.t}_{t.}, &
 G^{.t}_{z.}\!=\!
 \Big({\dot\rho\rho^\prime\rho^{\prime\prime}-
 \big(1+{\rho^\prime}^2\big)\dot\rho^\prime}\Big)
 \big/{{\rho\mathfrak{g}^2}}\!=\! T^{.t}_{z.}, \\
 \\
 G^{.z}_{t.}\!= \hskip .21truecm  \Big({ \dot\rho \rho^\prime
 \ddot\rho +\left(1-{\dot\rho}^2\right)\dot\rho^\prime}\Big)
 \big/{{\rho\mathfrak{g}^2}} \!=\! T^{.z}_{t.}, &
 G^{.z}_{z.}\!= \!\Big({\big(1+{\rho^\prime}^2\big)\ddot\rho-
 \dot\rho\rho^\prime \dot\rho^\prime}\Big)
 \big/{ {\rho\mathfrak{g}^2}}\!= \!T^{.z}_{z.},\!\!
\end{array}
 \!\!\right\}\hskip .5truecm
 \la{Enstn:a}\\
 \nonumber\\
 \begin{array}{ll} G^{.\phi}_{\phi.}= -
\Big({\ddot\rho\rho^{\prime\prime}-\big(\dot\rho^\prime\big)^2}\Big)
 \big/{{\mathfrak{g}^2}}= T^{.\phi}_{\phi.}\,,
 \hskip 4.truecm
\end{array}
\la{Enstn:b}
\een
\end{subequations}
where due to the above mentioned causality condition
$\mathfrak{g}=1-\dot\rho^2+{\rho^\prime}^2=g_{tt}g_{zz}-g_{tz}g_{zt}>0$
\footnote{The Riemannian
scalar curvature of the interval \eqref{ds2_1_2} is
$$R=-2\mathfrak{g}^{-1}\left(
\left(1-\dot\rho^2\right)\rho^{\prime\prime}
+2\dot\rho\rho^\prime \dot\rho^\prime
-\left(1+{\rho^\prime}^2\right)\ddot\rho +
\rho^2\big(\ddot\rho\rho^{\prime\prime}-(\dot\rho^\prime)^2\big)
\right).$$
Then the causality condition ensures the absence of curvature
singularities in the AxUs since\\
$\dot{\rho}^2 <1\,\,\,\Rightarrow\,\,\,\mathfrak{g}>0$.
}.

 The other four EEqs lead to zero components of the energy-momentum
 tensor of the matter sources:
\ben
T^{. \phi}_{t .}=T^{. \phi}_{z .}=T^{. t}_{\phi\,\,.}=T^{. z}_{\phi .}\equiv 0.
\la{zeroT}
\een
 These relations restrict the motion of the matter which creates
 a specific type of a universe with metric \eqref{ds2_1_2}.

 After some algebra one can write down Eqs. \eqref{Enstn:a} in a
 much simpler form:
\ben\la{MY}
\ddot\rho =-\rho\mathfrak{g}^2\, T^{zz}, \quad
\rho^{\prime\prime}=-\rho\mathfrak{g}^2\, T^{tt}, \quad
\dot\rho^\prime=\rho\mathfrak{g}^2\, T^{tz}.  \een

 Besides, one obtains from Eqs. \eqref{Enstn:a} the compatibility
 condition $T^{tz}=T^{zt}$ which is fulfilled by construction.
 The last Eq. \eqref{Enstn:b} in the Einstein system
 \eqref{Enstn:a,b} yields the constraint
\ben
 T^{\phi\phi}=\mathfrak{g}^2\left(T^{tt}T^{zz}-T^{tz}T^{zt}
 \right), \la{constrT_first} \een
 which shows that the component $T^{\phi\phi}$ is not an independent
 one and can be expressed in terms of the other nontrivial components
 of the energy-momentum tensor in AxU. This constraint can
 be represented in the form
 \ben\la{constrT}
 \det T=\left({ {T^{\phi\phi}}/{\mathfrak{g}}}\right)^2\geq 0 ,
 \een
 where $\det T=\det||T^{ij}||$ is the determinant of the
 contra-variant energy momentum tensors of matter in the
 $(2+1)$-dim AxUs, subject to the conditions, Eq. \eqref{zeroT}.
 The constraint \eqref{constrT} shows that the determinant
 $\det T\geq 0$ is a non-negative quantity. This is compatible with
 the properties of the energy-momentum tensor for physical matter
 in the $(2+1)$-dim AxUs, say for perfect fluid with standard
 eigenvalue $\epsilon\geq 0$ ($\epsilon$ being the density of
 energy) and two identical negative ones:  $-p\leq 0$ ($p$ being
 the pressure). Thus, relation \eqref{constrT} supports
 the compatibility of our models of AxUs with the standard physics
 \cite{Landau}.

 Note that Eqs. \eqref{MY} replace EEqs. \eqref{Enstn:a,b}
 and govern the dynamics of geometry of the AxUs. To obtain the whole
 dynamics of the universe, filled with some matter, one has to add
 the continuity equation
 \ben\la{MattEq}
 \nabla_iT^i_j=0,\quad i,j = 1,2,3.\een
 It is a well-known consequence of the EEqs. yielded by the
 restriction of the Bianchi identity on the Einstein tensor and
 presents the GR dynamical equations for matter in any
 $(2+1)$-dim universe. The third of the equations \eqref{MattEq}
 $\nabla_iT^i_3\equiv 0$ is identically fulfilled in the AxUs at
 hand.
 Hence, for AxUs we have a specific
 universally conserved vector quantity
\ben
T^i=T^i_j\xi^j,\quad \nabla_iT^i=0,
\la{conserved_vector}
 \een
 due to the axial symmetry which yields the obvious Killing vector
 $\boldsymbol{\xi}_\phi=\{\xi^t,\xi^z,\xi^\phi\}=\{0,0,1\}$ and
 conservation of the $z$-component of the angular momentum.

\section{The Solutions to the Einstein Equations for the Time Dependent AxU}

\subsection{The Vacuum Solutions of the Einstein Equations}

 The $(2+1)$-dim {\em vacuum} dynamical equations \eqref{MY} with
 zero right-hand-sides obey three simple solutions
 (related by Lorentz transformations):
\begin{enumerate}\label{sol_1_2}
   \item $\rho(t,z)=v_0(t-t_0)+\rho_0$, where $\rho_0\geq 0$ is an
 arbitrary constant, $v_0$ is the constant velocity of the expansion
 of the $1$-dim string all points of which are non-isolated DRP
 (described by the equation $\rho\equiv 0$ at the time instant
 $t=t_0-\rho_0/v_0$, if $0<|v_0|\leq 1$) on the surface of the
 cylinder of the radius $\rho(t)$ which is independent of the
 coordinate $z$;
  \item $\rho(t,z)=(z-z_0)\tan\alpha$, where $\alpha$ is the constant
 angle at the vertex of a static cone. Further on, the short notation
 $\sigma=\tan\alpha$ is used. The static isolated DRP is the point
 $z=z_0=\text{const}$;
  \item $\rho(t,z)=v_0(t-t_0)+\sigma(z-z_0)$, $v_0\neq 0$ is the
 velocity of a moving $2$-dim cone with the vertex angle
 $\alpha\in (0,\pi/2)$.
  Here an isolated running DRP $z^{drp}(t)=z_0-v(t-t_0)/\sigma$
 moves with constant velocity $v_0/\sigma\,.$ 
   \end{enumerate}

 As seen, in any case we
 have DRP of the EEqs. solutions,
 which are related to a reduction of the topological space-time
 dimension from $(2+1)$ to $(1+1)$, or even to $(1+0)$.

\subsection{The Solutions with a Positive $\mathbf{\Lambda}$ Term}

 In the case of {\em positive lambda} term $\Lambda=1/R^2> 0$, one
 has the only solution of the EEqs. $G^i_j=\Lambda \delta^i_j$ in
 the AxU:
$$\rho(t,z)=\sqrt{R^2-(z-z_0)^2}.$$
 It describes a $2$-dim static  spherical surface of
 constant radius $R$. This solution was briefly discussed in
 \cite{DVPF10}. On this sphere we have two isolated static DRPs:
 $z=z_0\pm R$ which are not singular points of the very surface.

\subsection{The Solutions for the $(2+1)$-dim Axial Universe Filled with Dust}

 It is clear that putting some matter content like "dust", perfect fluid, or different
 matter fields in the $(2+1)$-dim universe with variable axial geometry one
 can obtain much more sophisticated solutions of Einstein Eqs..
 Consider, for example, the case of this sort of a universe filled with dust. Then
\ben
T^{i j}= \mu(t,z)u^i(t,z)u^j(t,z),
\la{Tij}
\een
 where the standard notation was used \cite{Landau}.

 \subsubsection{The Solution to the gravitational field equations}
 As a result, from Eq. \eqref{constrT_first} one obtains $T^{\phi\phi}\equiv 0$ and
 the variable shape function $\rho(t,z)\geq 0$ has to be found according to
 Eq. \eqref{Enstn:b} by solving the well-known homogeneous Monge-Amp\`{e}re equation
 \cite{PZ} 
\ben
\ddot\rho\rho^{\prime\prime}-\left(\dot\rho^\prime\right)^2=0.
\la{MAE}
\een
Its general solution has the following implicit form in terms of two arbitrary functions $a(v)$ and $b(v)$:
\ben
\rho= t v +a(v)z+b(v),\quad t+a_{,v}(v)z +b_{,v}=0,
\la{GS_MA}
\een
where the comma denotes the corresponding partial differentiation.
From the second equation one has to obtain the function $v(t,z)=\dot\rho(t,z)$. This is possible if and only if
the following condition is fulfilled:
 \ben \la{cond_v}
 a_{,vv}(v)z +b_{,vv}\neq 0. \een
 After that one obtains the solution $\rho(t,z)$ from the first of
 the Eqs. \eqref{GS_MA}. In addition one gets the relations
 \ben\la{rel}
 \hskip -1.7truecm \rho^\prime=a(v),\quad v^\prime=\dot v a_{,v},\quad \ddot\rho=
 \dot v,\quad \rho^{\prime\prime}=v^\prime a_{,v}\quad
 \dot\rho^\prime=\dot va_{,v}\quad b(v)=\rho-t\dot\rho-z\rho^\prime,
 \een
 which reveal the meaning of the arbitrary functions in
 \eqref{GS_MA}. In particular, the function $b(v)$ describes the
 deviation of the shape function $\rho(t,z)$ from a homogeneous
 function of degree one.

 In the case $a_{,vv}(v)z +b_{,vv}\equiv 0$, Eq. \eqref{MAE} has a
 special solution
\ben\la{SingSol}
\rho(t,z)=v_0t+\rho_0(z), \quad |v_0|<1\,,\een
 $v_0$ being a constant velocity, not greater than the light
 velocity and $\rho_0(z)\geq 0$ being an arbitrary time-independent
 shape function.

It is not difficult also to obtain the general solution of the Cauchy problem.
Let $\rho_0(z)\geq 0$ and $\dot\rho_0(z)$ be the Cauchy data. Then using Eqs. \eqref{GS_MA} and  \eqref{rel}
one obtains
\begin{subequations}\la{Cauchy:a,b}
\ben
\rho(t,z)=\rho_0(\zeta)+(z-\zeta)\rho_{0,z}(\zeta)-t\dot\rho_0(\zeta), \la{Cauchy:a}\\
\hskip -2.5truecm \text{where $\zeta$ is defined by the equation}\nonumber \\
t\dot\rho_{0,z}(\zeta)= (z-\zeta)\rho_{0,z z}(\zeta)\quad \Rightarrow \quad \zeta=\zeta(t,z).\la{Cauchy:b}
\een
\end{subequations}

  \subsubsection{The Solution to the matter equations} Since the
 gravitational field dynamics is already known, the description of
 the dynamics of matter is a simple algebraic task. From Eqs.
 \eqref{MY}, relations \eqref{zeroT}, \eqref{Tij} and \eqref{rel},
 as well as taking into account the normalization condition
 $g_{ij}u^iu^j=1$ and assuming $\dot v=\ddot\rho\neq 0$, one obtains
\begin{subequations}\la{matt_sol:a,b}
\ben
u^t(t,z)&=&-{\frac{a_{,v}}{\eta(v)}}, \quad u^z(t,z)=
 {\frac{1}{\eta(v)}}, \quad  u^\phi(t,z)= 0, \la{matt_sol:a}\\
 \mu(t,z)&=& -\ddot\rho\left({\frac{\eta(v)}{1-v^2+a^2}}\right)^2
 \geq 0,\quad \Rightarrow \ddot\rho\leq 0, \la{matt_sol:b}
 \la{matter_sol}\een
\end{subequations}
 thus reaching a complete description of the motion of matter which
 builds the universes under consideration. Here
 $$\eta(v)=\sqrt{(1-v^2)(a_{,v})^2+v(a^2)_{,v}-a^2-1}$$
 must be real. Hence, the quantity under the square root should be
 nonnegative. The corresponding differential inequality can be
 represented in the form:
\ben\la{a_cond}
 a_{,v}\geq {\frac{\sqrt{1-v^2+a^2} - v a}{1-v^2}} \quad \text{for}
 \quad v\in (-1,1).\een
 This gives an additional restriction on the admissible functions
 $a(v)$:
\ben\la{a_restriction}
 2\,a(v)\geq (1-v)a(-1)-(1+v)/a(-1)\quad\text{for}\quad v\in (-1,1).
\een

 One can simplify the consideration of matter dynamics using the
 standard co-moving frame (where the matter is at rest, i.e.,
  $u^i=\delta^i_0$), but we shall skip here the details.

 \subsubsection{Dynamics of the Dimensional Reduction Points in
 the $(2+1)$-dim Axial Universe Filled with Dust}

 The zeros of the initial shape function $\rho_0(z)\geq 0$ are
 DRPs for the axial space geometry. According to Eqs.
 \eqref{Cauchy:a,b}, in general, these are moving DRPs.

 It is interesting to know whether it is possible to create
 additional DRPs which are not zeros of the initial shape function,
 or to annihilate some of the existing ones during the time
 evolution of our models.
 The following simple example shows that this is possible.

 Consider the Cauchy initial data $\rho_0(z)=
 {\tfrac1 2}z^2/R+r$, $\dot\rho_0(z)={\tfrac1 2}v_0 z^2/R^2,\,r,R>0$.
 Then from Eq. \eqref{Cauchy:a,b} one has $\rho_0(t,z)={\tfrac1 2}
 z^2/(R+v_0 t) +r$. Here, at the initial moment $t=0$ there is no
 real DRPs. Depending on the sign of the velocity constant $v_0$
 two such DRPs  $z^{drp}_{\pm}(t)=\pm\sqrt{2r(-R-v_0t)}$ appear or
 disappear at the instant $t=-R/v_0$. Hence, we have a typical
 bifurcation problem. Since under the change of the corresponding
 bifurcation parameter the simple real roots of analytic functions
 occur in pairs, or disappear in pairs, this is also true for the
 DRPs in our problem assuming an analytical character of the Cauchy
 data.

 One can point out several quite general examples of the $(2+1)$-dim
 AxUs in which the number of the DRPs (finite, or even infinite one)
 is constant during the time evolution, see the Appendix A. It is
 possible that there exists an infinite sequence of DRPs which has
 a finite limiting point. Hence, the structure and the dynamics of
 the DRPs of the universe may be quite complicated.

 \section{The Solutions of the KFGEq on the $\mathbf{(2+1)}$-dim
   AxUs which are the EEqs solutions}

 The test particles and fields of any spin in the (2+1)-dim AxUs
 have a common property. Due to the axial symmetry the $z$-component
 of their angular momentum is a constant of motion. For the KFGEq
\begin{equation}\label{KGE}
\Box\varphi-M^2\varphi=0,\qquad \Box=-\frac{1}{\sqrt{|g|}}\,
\partial_\mu(\sqrt{|g|}g^{\mu\nu}\,\partial_\nu)\end{equation}
 this means that one can separate the angular part of the field
 by the ansatz $\varphi(t,z,\phi)=f_m(t,z)e^{i m \phi}$, where the
 azimuthal number $m=0,\pm 1, \pm 2, \dots$ is an integer.

\subsection{The Solutions of KFGEq on the $\mathbf{(2+1)}$-dim AxUs
   which are the EEqs Vacuum solutions}

 1. For simplicity, we write down the first vacuum solution of
 Sect.IV.A in the form $\rho(t,z)=vt$ ($v=\text{const}$) which shows
 that it describes a cylinder with a radius independent of the
 variable $z$. This cylinder collapses to a thread
 with a zero radius at instant $t=0$ (for $v<0$), or vice-versa,
 thread  expands to a cylinder with increasing radius
 $\rho(t,z)=v t$ (if $v>0$). The corresponding KFGEq reads
 $$
 -{\tfrac1 {1-v^2}} {\tfrac1 {|t|}} \partial_t \big(|t|\partial_t
  \varphi\big)+\partial_z^2\varphi +
{\tfrac {1}{v^2 t^2}} \partial_\phi^2 \varphi - M^2 \varphi=0.$$
 After separation of variables one obtains its solutions in the
 interval \ $t\in (+0,+\infty)$ \ 
 \ben \la{phi_t_z_1}
 \varphi_1(t,z,\phi)= J_\nu\left(\omega_p t\right) e^{ip_zz}
 e^{im\phi},\quad \varphi_2(t,z,\phi)= Y_\nu\left(\omega_p t\right)
  e^{ip_zz} e^{im\phi},\\ \nu=i\sqrt{1-v^2}\,|m|/v, \quad\quad
 \omega_p=\sqrt{(1-v^2)(p_z^2+M^2)},\nonumber \een
 where $p_z$ is the $z$-component of the momentum of the Klein-Fock-Gordon
 field. Since the index $\nu$ of the Bessel functions $J_\nu$ and
 $Y_\nu$ is a purely imaginary number, in the limit $t\to +0$, when
 the space becomes $1$-dim, both solutions oscillate infinitely
 many times remaining limited in the amplitude. The solutions can
 not be continued  directly through the DRP $t=0$ which is an
 infinite-branching point on the real axis $t\,.$

 2. In the case 2 of Section IV. A, the solution describes a static
 cone $\rho(t,z)=\sigma(z-z_0)$. The only static DRP lies on the
 $Oz$ axis at the point $z=z_0\,.$ The different solutions of the
 KFGEq on this type of cone were found in \cite{PFDV10}. We
 give them for comparison with other solutions considered here
 \ben\la{phi_t_z_2} \varphi_1(t,z,\phi)= e^{-i\omega t}
  J_\nu\left(k_c\,z\right) e^{im\phi},\quad \varphi_2(t,z,\phi)=
  e^{-i\omega t} Y_\nu\left(k_c \,z\right)e^{im\phi},\\
  \nu=|m|/\sin\alpha,\qquad 
 k_{c}=\sqrt{\omega^2-M^2}/\cos\alpha,\nonumber \een
 where $\omega$ is in general complex frequency with a positive
 imaginary part. As a result, the first solution
 $\varphi_1(t,z,\phi)$ vanishes at the static DRP $z=0$, while the
 second one $\varphi_2(t,z,\phi)$ diverges.

 In \cite{PFDV10} one can also find a detailed description of
 highly nontrivial excitations of the Klein-Fock-Gordon field on a
 continuous manifold built of the parts of two static axial
 surfaces of type 1, given in Section IV.A, but now with $v=0$ and
 different constant radii $\rho_{0,1}=r\,,\,\,\rho_{0,2}=R>\,,$
 connected by the corresponding part of the static cone.

 3. Turn now to the third  case described in Section IV.A. The
 solution $\rho(t,z)=\sigma z +v t$ ($\sigma ,v\neq 0$)
 represents a moving $2$-dim cone. It has a moving DRP \
 $z^{drp}(t)=-v t/\sigma $.

 After separation of the variables in the KFGEq by specific ansatz
 $\varphi(t,z,\phi)=F(\sigma z+vt)G(vz+\sigma t)e^{i m \phi}$
 one obtains two independent solutions:
 \ben\la{phi_t_z_3}
 \varphi_1(t,z,\phi)= e^{-ia{\frac{(vz+\sigma t)}{\sqrt{\sigma^2-
 v^2}}}}J_\nu\big(\varkappa_a (\sigma z+vt)\big) e^{im\phi},\nonumber\\
 \varphi_2(t,z,\phi)= e^{-ia{\frac{(vz+\sigma t)}{\sqrt{\sigma^2-
 v^2}}}} Y_\nu\big(\varkappa_a (\sigma z+vt)\big) e^{im\phi},\\
 \nu=|m|\sqrt{\tfrac{1-v^2+\sigma^2}{\sigma^2-v^2}},\,\,
 \varkappa_a=\sqrt{{\tfrac{1-v^2+\sigma^2}{\sigma^2-
 v^2}}\left(a^2-M^2\right)}.\nonumber\een
 Here, the separation constant $a$ plays the role of a spectral
 parameter. To obtain Eqs.\eqref{phi_t_z_1} from Eqs.
 \eqref{phi_t_z_3}, one has to put $\sigma =0$ and $a=ip_z$. For
 obtaining Eqs. \eqref{phi_t_z_2}  one has to put $v=0$ and
 $a=\omega\,.$

 Note that in the vicinity of the moving DRP $z^{drp}(t)=-
 v t/\sigma $ the solutions \eqref{phi_t_z_3} of the KFGEq have a
 different behavior depending on the values of the constants $v$
 and $\sigma = \tan\alpha$.

 \begin{enumerate}\label{sol_phi_t_z_3}
  \item If $v\cot\alpha \geq 1$, then $\nu$ is imaginary, the
 DRP moves with superluminal velocity and both the solutions
 \eqref{phi_t_z_3} are bounded in its vicinity but make an infinite
 number of oscillations approaching this point. For comparison,
 see the etalon case 1, i.e.,  Eqs. \eqref{phi_t_z_1} when
 $v\cot\alpha =\infty$.
  \item If $v\cot\alpha \leq 1$, then $\nu>0$ is real, the DRP
 moves with subluminal velocity and the solution
 $\varphi_1(t,z,\phi)\to 0$ but $\varphi_1(t,z,\phi)\to \infty$
 in the vicinity of the singular point. For comparison, see the
 etalon case 2, i.e., Eqs. \eqref{phi_t_z_2} when $v\cot\alpha =0$.
 \end{enumerate}

 \subsection{The Solutions of the KFGEq on the $\mathbf{(2+1)}$-dim
   AxU with  a Positive \ $\Lambda$ \ Term}
 In this case (see Section IV. B.), the standard separation of the
 variables in the KFGEq leads to the following two solutions in terms
 of the Legendre functions\footnote{We use here the \textsc{maple}
 notation for the Legendre functions assuming branch cuts on the real
 semi-axes $(-\infty,-1)$ and $(1,+\infty)$. This choice of the branch cuts
 is most convenient for our problem defined on the real interval $[-1,1]$
 and differs from the standard definitions of the Legendre functions
 in \cite{SpecialFunctionsA} and \cite{SpecialFunctionsB}.
  For each of the two admissible values of $\nu$:
  $\nu=\pm\,\sqrt{1/4+R^2\left(\omega^2-M^2\right)}-1/2$
 one obtains two solutions \eqref{LegendreP} which are well defined
 on the interval $z\in (-R, R)$ and are in general linearly
 independent.}:
\ben
 \varphi_{1,2}(t,z,\phi)&=& e^{-i\omega t}\text{LegendreP}
 \left(\nu,|m|, \pm z/R\right) e^{im\phi}.
\la{LegendreP}
\een
 Then $\varphi_{1}(t,z,\phi)$ is regular at the point $z=R$ and
 singular at the point $z=-R$, while $\varphi_{2}(t,z,\phi)$ is
 regular at the point $z=-R$ and singular at the point $z=R$.

 The space of our static universe with the $\Lambda$ term is a
 closed $2$-dim spherical surface. One obtains an infinite series
 of everywhere regular solutions of KFGEq $\varphi(t,z,\phi;n,m)$
 which have a discrete spectrum with real frequencies:
\ben\la{spectrum_121}
 \omega_{n,m}^\pm=\sqrt{M^2+(n \pm|m|)(n \pm|m|+1)/R^2},\quad
 n=0,1,2,\dots \een
 imposing the requirement for linear dependence of the solutions
 $\varphi_{1}(t,z,\phi)$ and $\varphi_{2}(t,z,\phi)$, defined
 by  Eq. \eqref{LegendreP}. This assigns integer values to the
 parameter $\nu$ in Eq. \eqref{LegendreP} and brings us to the
 associated Legendre polynomials.

\subsection{Some Solutions of the KFGEq on the $\mathbf{(2+1)}$-dim Axial
 Universe Filled with Dust}
 Consider for example solution \eqref{sol:a} which presents a
 moving wave $\rho(t,z)=f(\sigma z+tv)$. After some algebra one
 separates the variables in the corresponding KFGEq by using again
 the ansatz
$\varphi(t,z,\phi)=F(\sigma z+vt)G(vz+\sigma t)e^{i m \phi}$
 and obtains
 $G(vz+\sigma t)=\exp(-ia(vz+\sigma t)/\sqrt{\sigma^2-v^2})$.
 Now the function $F(x)$ has to be a solution of the following
 ODE defined by the function $f(x)$:
\ben\la{F_x}\hskip -2.3truecm
F^{\prime\prime}+
 {\frac{ff^\prime\left((\sigma^2-v^2)(f^\prime/f)^\prime-1\right)}
 {1+(\sigma^2-v^2)(f^\prime)^2}}F^\prime+{\frac {1+(\sigma^2-v^2)
 (f^\prime)^2}{\sigma^2-v^2}}\left(a^2-M^2-{\frac{m^2}{f^2}}\right)F=0.
\een
 Here the prime denotes differentiation with respect to the variable
 $x=\sigma z+vt$. For some specific functions $f(x)$ this equation
 has two independent solutions $F_{1,2}(x)$ and one arrives at the
 KFGEq solutions in the form
 \ben
 \varphi_{1,2}(t,z,\phi)= exp\left\{-i{\tfrac{(vz+\sigma t)}
 {\sqrt{\sigma^2-v^2}}}\right\} F_{1,2}(\sigma z+vt)\,e^{im\phi}.
\la{phi_t_z_4}
 \een
 For the simple case $f(x)=x$ this gives the already obtained
 vacuum result \eqref{phi_t_z_3}. For $f(x)=\sqrt{x}$ one obtains
 exact solutions of Eq. \eqref{phi_t_z_4} in terms of the
 confluent Heun function:
 \ben\la{Heun}\hskip -1.3truecm
 F_{1,2}(x)=x^{\pm\frac m 2}exp\left\{i\sqrt{\tfrac{a^2-M^2}
 {\sigma^2-v^2}}x\right\}\text{HeunC}\left(\alpha,\pm m,-
  {\tfrac 3 2},\delta,{\tfrac {3} 4}-\delta,-
  {\frac {4x}{\sigma^2-v^2}}\right)\,,\een
 where $\alpha=i\sqrt{(a^2-M^2)(\sigma^2-v^2)}$ and
 $\delta={\tfrac {m^2} 4}-{\tfrac1{16}}\left(\sigma^2-v^2\right)
 \left(a^2-M^2\right).$ Since
 $\text{HeunC}(\alpha,\beta,\gamma,\delta,\eta,0)=1$, around the
 moving DRP $z^{drp}=-vt/\sigma$ the behavior of the KFGEq solutions
 is
 \ben \la{}
 \varphi_{1,2}\sim (\sigma z+vt)^{\pm\frac m 2}
 \exp\left(i\left[{\tfrac {\sigma\sqrt{a^2-M^2}-a v}
 {\sqrt{\sigma^2-v^2}}}\,z + {\tfrac {v\sqrt{a^2-
 M^2}-a \sigma}{\sqrt{\sigma^2-v^2}}}\,t \right]\right). \een

 \section{Summary and Outlook}
    Our main results are: \smallskip

 1. A special kind of the $(2+1)$-dim toy models of time dependent
 universes with axial space ("axial universes") is introduced and
 considered in detail. Their variable
 geometry is defined by a single time dependent shape function
 $\rho(t,z)\geq 0$. It is shown that these models allow the study
 of topological-dimension reduction phenomena since the zeros of
 the shape function define the dimensional reduction points;\smallskip

 2. The time dependent shape function $\rho(t,z)$ of the axial
 universes (AxUs) is determined by the solution of the Einstein
 equations (EEqs) for various  energy-momentum tensors of
 matter. The exact solutions for the vacuum AxUs, AxUs filled
 with the $\Lambda$ term, as well as the exact general solution of the
 EEqs for the AxUs filled with dust are found. In the last case, we
 reduce the EEqs solution to solutions of the homogeneous
 Monge-Amp\`{e}re equation and obtain the general solution of the
 Cauchy problem for time dependent AxUs, as well as three
 independent classes of solution each of which involves an arbitrary
 function. It is shown that some of the previously considered
 static AxUs are solutions to the EEqs;\smallskip

 3. The dynamics of the dimensional reduction points in the
 time dependent AxUs is studied. It is found that these points can
 emerge and disappear in a real domain. Their dynamics is described
 in three essentially different cases;\smallskip

 4. The spreading of test particles in variable geometry, including
 reduction of topological dimension is studied. For this purpose
 the exact solutions for test particles described by the
 Klein-Fock-Gordon eq. in different time-dependent AxUs are presented
 in terms of special functions. The behavior of test particles in
 the vicinity of the dimensional reduction points is described
 explicitly \smallskip

These results allow us to express a hope for a possible further development:
 \smallskip

 $\bullet$ The time dependent AxUs with variable space-time geometry could
  give one a simple basis for commenting the real situation concerning
  the C, P, and also T properties of the particles.\smallskip

 $\bullet$ The considered models of time-dependent AxUs inspire an intriguing
 idea: to treat the very Big Bang as a transition from the
 pre-Big-Bang-Universe with a lower topological dimension ($d=1$, or
 $d=2$): (or as a sequence of such transitions), to the present-day
 space with $d=3$. For this purpose a generalization of the model,
 described in \cite{Fiziev2010}, is needed.\smallskip

At present a quite intensive discussion of a different kind of reduction of space dimensions
related with the theory of gravity and particle physics takes place in the literature
(See the Appendix B and the references therein.).
Here we outline our answer to a very important question raised in \cite{Visser}:
\begin{quote}
 {\it"... there is no reason whatsoever for the theory ...
 to be close to 2+1 dimensional GR in the UV. Clearly, if this is
 to be a viable gravity theory it should resemble 4-dimensional
 GR at low energies. In fact, it is difficult to imagine how a
 theory that has this latter property can have the former property
 as well...".}  \end{quote}

The same question also applies to our approach to dimensional reduction
although it differs significantly from the others.

First of all, in our approach the dimensional reduction is a dynamical consequence of EEqs in GR.
As a result, one may expect that in the smaller dimensions an analogous theory will be valid
in the same manner as in the well-known Kaluza-Klein theories.
This problem needs careful justification, especially for more realistic transitions between
$(3+1)$-dim, $(2+1)$-dim, and $(1+1)$-dim geometries,
starting or ending with usual $(3+1)$-dim GR with a different kind of matter content of the Universe.
The present article is to be considered only as a first step in this direction,
based on the very simplified AxUs toy models.

 Using the experience in the Kaluza-Klein-like theories (see, for
 example \cite{KK}, and the references therein), we can shed more light
 on the above consideration.
 Indeed, it is well known that after topological-dimension
 reduction in the modern Kaluza-Klein theories one remains with
 usual GR and some set of additional fields of various spins.
 This set is determined by the theory in higher dimensions. Hence,
 if we intend to reproduce our observable world, starting from a
 {\it pre-Universe with lower dimension}, the last has to be
 equipped with a definite set of primordial fields, precisely
 adjusted to reconstructing our 4-dim world.
 \smallskip

 $\bullet$ Therefore, it is reasonable to continue the development of
 this type of models by solving the inverse problem: to
 reconstruct the physics in the lower dimensional pre-Universe
 which allows one to reproduce the known physics in our real
 world. One of the important methods to relate the observable
 properties with the variable geometry is to use the
 "fingerprints" of the shape of the junction domain between parts
 of the pre-Universe with different topological dimensions on
 the spectra of the observed particles \cite{PFDV10}.

 \medskip
  Besides the use of lower space-time dimensions in the place of
 higher ones, the main difference between the Kaluza-Klein theories
 with their further development  (like superstring models) and our
 approach is that in the latter, instead of fixing very small
 compactification radii needed to make the introduced
 additional higher dimensions unobservable, we use the shape function\footnote{Or
 a set of shape functions, see
 \cite{Fiziev2010}.} $\rho(t,z)\geq 0$ (solution of the dynamic
 equations) which play the role of a variable "compactification radius".
 If $\rho(t,z)\to\infty$, we have a flat space. In the opposite case,
 when $\rho(t,z)\to 0$ the topological dimension of
 the space-time reduces, as seen from the results of the present
 article\footnote{Thus, our model of topological dimensional
 reduction is fundamentally different from models that use fractal,
 spectral or other possible concepts of dimensionality of
 space-time, see the Appendix B.}.

\section*{Acknowledgments}
 It is a pleasure to thank \ George Alekseev, Irina Aref'eva, Valeri
 Rubakov and Oleg Teryaev for useful discussions and the strong
 pulses for movement in the right direction of investigations.
 The conversations with  Vasily Neznamov were very stimulating, too.

 The research has been partially supported by the Bulgarian
 National Scientific Fund under contracts DO-1-872, DO-1-895,
 DO-02-136, by the Sofia University Scientific Fund, contract
 185/26.04.2010 as well as by the Russian Presidential grant,
 Scientific School--3810.2010.2 and by RFBR grant 11-01-00182,

We are grateful to the unknown referees for the useful suggestions and especially
for pointing out the references \cite{Visser09} and \cite{Zanelli},
as well as to Lev Okun for drawing our attention
to the references \cite{Klein,Fock_a,Fock_b,Gordon}.

\appendix

\section{Examples of time evolution of the set of dimensional reduction points in
the $\mathbf{(2+1)}$-dim  axial universe, filled with dust}

One can shed some additional light on the evolution of  DRPs  considering the following three
different types of solutions $\rho(t,z)$ in which only one arbitrary function $f(x)$ is involved \cite{PZ}:
\begin{subequations}\la{sol:a,b,c}
\ben
\rho(t,z)&=&f(\sigma_0 z-v_0 t)+\sigma_{1} z- v_1 t+\rho_1\la{sol:a},\\
\rho(t,z)&=&(\sigma_0 z - v_0 t + \rho_0)
f\left({\frac {\sigma_1 z-v_1 t+\rho_1}{\sigma_0 z - v_0 t + \rho_0}}\right)+\sigma_{2} z - v_{2} t+\rho_{2},\la{sol:b}\\
\rho(t,z)&=&(\sigma_0 z-v_0 t)f\left({ {v t}/ z}\right)+\sigma_{1} z - v_{1} t+\rho_{1},\la{sol:c}
\een
\end{subequations}
with arbitrary constants ${\sigma_0, \sigma_1, \sigma_2, v_0, v_1, v_2, \rho_{0}, \rho_{1}, \rho_{2}}$.

Suppose $x_{i=1,2,\dots}$ to be the zeros of the corresponding function $f(x)$
(which for this issue is assumed to be a {\em bounded} function) and consider three cases:

\begin{enumerate}\label{singularities}
  \item Running waves of type \eqref{sol:a} with  the equation
$\rho(t,z)=f(\sigma_0 z-v_0 t)$  (where $\sigma_0, v_0\neq 0$).
Then, the moving DRPs are $$z^{drp}_{i}(t)=(v_0 t+x_i)/\sigma_0,\quad i=1,2,\dots.$$
The distance between the different DRPs remains constant during the time evolution.
  \item Solutions of type \eqref{sol:b} with  the equation
$\rho(t,z)=(\sigma_0 (z-z_0) - v_0 t )f\left({\frac {\sigma_1 z-v_1 t+\rho_1}{\sigma_0 (z-z_0) - v_0 t }}\right)$
(where $\sigma_{0,1}, v_{0,1}\neq 0$).
Now we have DRP related with the first factor:
$z^{drp}_{0}(t)=z_0-{ {v_0}t/ {\sigma_0}}$  and additional DRPs
$$z^{drp}_{i}(t)=z_0+{\frac{v_1-v_0 x_i}{\sigma_1-\sigma_0 x_i}}t-{\frac{\rho_1-z_0 \sigma_1}{\sigma_1-\sigma_0 x_i}},\quad i=1,2,\dots.$$

Since different DRPs move with different constant velocities, starting from a different
initial position, their ordering may change, depending on the roots $x_i$.
During the time evolution some pairs of DRPs may coalesce.
Indeed, the relative velocity between the points $z^{drp}_{i}(t)$ and $z^{drp}_{j}(t)$ is constant:
$$v_{ij}={\frac {(x_i-x_j)(\sigma_0 v_1-\sigma_1v_0)}{(\sigma_1-x_i\sigma_0)(\sigma_1-x_j\sigma_0)}}.$$
Hence, if there are DRPs with $v_{ij}<0$, they coalesce
and after that go away from each other.
  \item Solutions of type \eqref{sol:c} with  the equation $\rho(t,z)=(\sigma_0 z-v_0 t)f\left({ {vt}/ z}\right)$
  (with $\sigma_0, v_0, v\neq 0$).
 Related with the first factor is the DRP $z^{drp}_{0}(t)=- {v_0}t/ {\sigma_0}$.
 The other ones are
 $$z^{drp}_{i}(t)=vt/x_i, \quad i=1,2,\dots.$$
 In this case all DRPs start from the common origin
 $z=0$ at the time instant $t=0$ and move with different velocities.
\end{enumerate}

In all three cases the number of DRPs is constant during the time evolution of the universe.

\section{Remarks on some other approaches to dimensional reduction}

In the literature (See, for example, \cite{Ambjorn98,Ambjorn01,Ambjorn04,Ambjorn05,Horava,Anchor11,Landsberg,Stojkovich1,Stojkovich2}
and the references therein.) one can find quite different attempts to consider space-time with variable dimension.

It seems that for the first time the idea of lowering the space dimension at short distances appeared
in the models with causal dynamical triangulations (CDT) \cite{Ambjorn98,Ambjorn01,Ambjorn04,Ambjorn05}
in which the geometry emerges as the sum of all possible triangulations.
There, a specific "spectral dimension" was introduced as a probe of space dimension in CDT.
The spectral dimension need not be an integer and furthermore is scale-dependent.
It can be thought of as the effective dimension as probed by a fictitious diffusion process.

In \cite{Horava}, a good agreement was found between the spectral dimensions of CDT and the completely different
Horava-Lifshitz (HL) model of gravity based on a specific violation of Lorentz invariance at small distances.

Due to the articles \cite{Visser11a, Visser11b} now we know that
the spectral dimension, as opposed to the topological dimension, actually
is related with the kinematics of the fictitious diffusion process (as well as of other possible dynamical processes)
in the region where space-time curvature is small and the manifold is flat to a good approximation.
Hence, the spectral dimension is not necessarily intrinsically geometric.
It is ultimately equivalent to a dispersion relation for the differential operator which describes
the dynamical process under consideration.

As we see, the above two approaches to dimensional reduction are based on completely different physical properties and
differ essentially from our approach which is based on the standard topological dimension of the space-time and considers
physical processes on manifolds with {\em variable topological dimension}. Thus, the simultaneous use
of the term "dimensional reduction" both in CDT or HL and in our model, is misleading.

Another class of models which make an attempt to use space-time with variable dimension
are described in \cite{Anchor11,Landsberg,Stojkovich1,Stojkovich2} and references therein.
Actually, in these articles we have no well defined model
but only some general ideas, inspired by some yet not explained observational data from processes
with cosmic ray particles detected in the Pamir mountains:
"an alignment of the main energy fluxes along a straight line in a target plane".
This observation provokes the authors for a search for a layered structure of space.

At the beginning, the authors try to construct some crystal-like structure of the space
("ordered string/brane lattice"):
The space consists of $1$-dim pieces at very small distances,
joined in $2$-dim layers at larger distances.
In its turn the $2$-dim layers are joined in $3$-dim cells at usual scales
and then the last are joined in $4$-dim structure at distances
bigger than the size of the observable Universe.

A somewhat different picture of the Universe, guided by the same motivation,
is a folded string with folding given by the fundamental quantization scale $L_1$ at very high energy.
It then folds and interweaves forming a $2$-dim structure with the
characteristic scale $L_2>>L_1$ which in turn folds to make a $3$-dim structure at scale $L_3>>L_2$, etc.
It remains unclear which definition of dimension is used in the concept of
"vanishing dimensions". Nevertheless, we see that such concept is completely different from
the dimensional reduction scheme developed in the present article,
based on the topological dimension of manifolds and EEqs.

Using the above quite undefined general ideas, the authors of \cite{Stojkovich1}
propose a physical speculation according to which
one can detect the vanishing dimensions via primordial gravitational wave astronomy:
If the Universe was indeed $(2+1)$-dim at some earlier epoch,
according to the authors, we will not see gravitational waves from this epoch.

The last conclusion is strongly criticized in \cite{Visser}.
We can support this critic using another qualitative argument:
According to the already discussed analogy with the Kaluza-Klein models,
in the $(1+1)$-dim and/or $(2+1)$-dim phase of the Universe we have to introduce primordial fields
from which after the transition to $(3+1)$-dim phase will be constructed the usual GR gravitational waves.
In our model the low-dimensional pre-Universe is not empty. As a result, in the $(3+1)$-dim phase of the Universe
we can see the signals of the primordial fields in the form of standard $(3+1)$-dim GR gravitational waves, since,
according to the results of \cite{PFDV10}, the penetration of signals from low the dimensional part of space
into the higher dimensional one is possible.

It is important to stress that as well as in the Kaluza-Klein models, in our design of dimensional reduction the
numerical-field-degrees-of-freedom of the theory are preserved during the transition from one to another
space-time dimension.
The only thing that changes is the grouping of these field-degrees-of-freedom in different multiplets,
because of the different symmetry groups in the tangent foliations of the parts
with different dimension of the space-time manifold.


\end{document}